\documentclass{ws-jai}
\usepackage{graphicx}
\usepackage{amsmath}
\usepackage{nicefrac}
\usepackage{algorithm}
\usepackage{algorithmic}

\begin{document}

\catchline{}{}{}{}{} 

\markboth{Alessio Magro}{GPU-Powered Coherent Beamforming}

\title{GPU-Powered Coherent Beamforming}

\author{A. ~Magro$^{1,2}$, K. ~Zarb ~Adami$^{1,2,3}$ and J. ~Hickish$^{4}$}

\address{
$^1$Institue of Space Sciences and Astronomy (ISSA), University of Malta, Msida 
MSD 2080, Malta \\
$^2$Department of Physics, University of Malta, Msida, MSD 2080, Malta \\
$^3$Department of Physics, University of Oxford, Denys Wilkinson Building, 
Keble Road, Oxford OX1 3RH. UK \\
$^4$Astrophysics Group, Cavendish Laboratory, University of Cambridge, JJ 
Thomson Avenue, Cambridge CB3 0HE, UK 
}

\maketitle

\begin{history}
\received{(to be inserted by publisher)};
\revised{(to be inserted by publisher)};
\accepted{(to be inserted by publisher)};
\end{history}

\begin{abstract}

GPU-based beamforming is a relatively unexplored area in radio astronomy, 
possibly due to the assumption that any such system will be severely limited by 
the PCIe bandwidth required to transfer data to the GPU. We have developed a 
CUDA-based GPU implementation of a coherent beamformer, specifically designed
and optimised for deployment at the BEST-2 array which can generate an 
arbitrary number of synthesized beams for a wide range of parameters. It 
achieves $\sim$1.3 TFLOPs on an NVIDIA Tesla K20, approximately 10x faster than 
an optimised, multithreaded CPU implementation. This kernel has been integrated 
into two real-time, GPU-based time-domain software pipelines deployed at the 
BEST-2 array in Medicina: a standalone beamforming pipeline and a transient 
detection pipeline. We present performance benchmarks for the beamforming 
kernel as well as the transient detection pipeline with beamforming 
capabilities as well as results of test observation. 

\end{abstract}

\section{Introduction}
\label{introductionSection}

In many signal processing applications an array of sensors, or antennas, need 
to be combined together such that the entire system can be used as a single 
entity. This is the process of beamforming, where the signals from each 
element arriving at different times are aligned in time/phase so that they can 
add coherently. 

In an array with $N$ elements a radiation wavefront originating from a 
particular direction will reach element $n$ at time $t_n$. Squaring and adding 
the signals of all the elements together is the process of incoherent 
beamforming. The maximum amplitude is achieved when the signals originate from 
a 
source perpendicular to the array, where they are highly correlated and add 
constructively. Alternatively, if the signals originate from a 
non-perpendicular 
direction they will arrive at different times, will be less correlated and 
result in a lower output amplitude. Coherent beamforming relies on the fact 
that 
for a given array configuration the relative delays between arrival times 
$t_{0..N-1}$ are a function of the direction of propagation of the incident 
wave. Artificial delays or phase shifts, in the time and frequency domain 
respectively, can be applied to the received signals from each antenna, causing 
them to add constructively when element signals are summed.

After summing to form a beam towards the incident radiation with wavevector 
$k_0$ the response $B(k)$ of the beam to radiation with other wavevector 
directions $k(\theta, \phi)$ is given by summing the signal contributions from 
individual antennas. If each individual antenna in an array has a response to 
radiation given by $A_n(k)$, this is given by
\begin{equation}
\label{beamResponseEquation}
 B(k)=\sum\limits_{n=0}^{N-1}A_n(k)e^{i(k-k_0)\cdot r_n}
\end{equation}
where $r_n$ is the position vector of antenna $n$. This is equivalent to the 
Fourier transform of the contributions from receiving elements weighted by 
their individual antenna responses. In cases where the antennas have very 
similar response functions Equation \ref{beamResponseEquation} reduces to the 
Fourier transform of the antenna distribution modified by the response pattern 
of each antenna. This technique was implemented for the BEST-2 array by 
\cite{Foster2014}. In this case, the Field-of-View (FoV) of the synthesised 
beam is approximately given by \nicefrac{$\lambda$}{$r_{\text{max}}$}, where 
$r_{\text{max}}$ is the maximum separation of the antennas used to form the 
beam.

Since prior to beamforming each individual antenna in an array sees a portion 
of the sky determined by its response function $A_n(k)$, it is possible to 
form useful beams in any direction where $A_n$ is significantly non-zero. 
It is also possible to generate multiple beams by making copies of the input 
antenna signals and adding them with different phase weightings, thereby 
increasing the coverage of an array's full FoV. Choosing 
the number of beams to form amounts to balancing the FoV observed by an array 
(which increases linearly with each beam added) with the amount of signal 
processing. Each beam can be treated as if it is a signal from a single 
antenna with response equal to the beam response, and so can be fed directly to 
backend detectors and used for time-domain observations, such as cosmic 
transient event observations, space debris detectors and transient surveys. The 
latter generally require a wide FoV and high sensitivity, however achieving 
this 
for larger telescopes can be prohibitively expensive in signal processing cost, 
especially for Discrete Fourier Transform (DFT) beamforming, since the number 
of required beams goes as $\left(D/d\right)^2$, where $D$ is the physical extent 
of the array and $d$ is the size of an individual element. Alternative 
strategies include sacrificing sensitivity to achieve maximal FoV using 
incoherent beamforming, or partitioning the array into multiple phased arrays 
and processing them independently. \cite{Colegate2011} describe these strategies 
with respect to transient searching for SKA$_1$. Signal processing cost can 
also be alleviated by using the regular layout of a telescope to remove 
redundancy in the beamforming process, for example as described by 
\cite{Tegmark2010}. This has been implemented in the digital backend deployed 
at the BEST-2 array \cite{Foster2014}, however will not be further investigated 
in this paper. 

In this paper we investigate the applicability of Graphics Processing Units 
(GPUs) for accelerating 
beamforming. We present an optimised implementation of a coherent beamformer in 
Section \ref{beamformingSection}, which is appropriately analysed and 
benchmarked. We also present a direct comparison with an optimsed 
Central Processing Unit (CPU)-version of 
the same algorithm. In Section \ref{best2Section} we present a beamforming 
system deployed at the BEST-2 array in Medicina, Italy, whilst in Section 
\ref{transientSection} we discuss the integration of this system with the the 
real-time transient detection system developed by \cite{Magro2013}.

\subsection{The BEST-2 Array and Digital Backend}
\label{best2Section}

The Basic Element for SKA Training II (BEST-2) \cite{Montebugnoli2009} is a 
subset of the Northern Cross cylindrical array at the Medicina observatory near 
Bologna, Italy. The array is composed of eight East-West oriented cylinderical 
concentrators, each with 64 dipole receivers spaced such that the cylindrical 
focal line is critically sampled at 408 MHz. Signals from 64 dipoles are 
combined in groups of 16 using analogue circuitry, resulting in four analogue 
channels per cylinder. This results in a total of 32 effective receiving 
elements positioned regularly on a 4 x 8 grid. The digital backend, developed 
and deployed by \cite{Foster2014}, is based on three 
ROACH\footnote{Reconfigurable Open Architecture Computing Hardware - 
https://casper.berkeley.edu/wiki/ROACH} processing boards and a number of 
software-based processing nodes. The three ROACH boards are referred to as the:
\begin{description}
 \item {\bf `F'-engine} for frequency transform, which is responsible for 
digitisation, channelisation of the processed 20 MHz bandwidth into 1024 
frequency subbands, and transmission of coarsely quantised antenna signals to 
downstream processing nodes using a custom 64ADCx64-12 
board\footnote{https://casper.berkeley.edu/wiki/64ADCx64-12}. A 4-tap 
Hann-windowed polyphase filter bank is used for channelisation.
\item {\bf `S'-engine} for spatial transform, which is responsible for 
formation of electric-field and total power beams on the sky by spatial Fourier 
transform and SPEAD\footnote{Streaming Protocol for Exchanging Astronomical 
Data 
 - https://casper.berkeley.edu/wiki/SPEAD} packetisation for 10 GbE streaming 
to 
the time domain 
processing server. The current implementation allows for the arbitrary 
selection 
of 8 beams from the generated 128-beam grid for output as 16-bit complex-valued 
words.
 \item {\bf `X'-engine} for cross-multiplication, which performs cross 
multiplication and accumulation of antenna signals as well as SPEAD 
packetisation of the generated visibility matrices and streaming to the 
visibility storage server for offline imaging. This is also used to calibrate 
the S-engine, such that F-engine streams are equalised and gain calibrated. 
\end{description}

Downstream processing of time-domain beam data is accomplished by using a 
Linux-based server which hosts 2 NVIDIA GPUs running the transient detection 
pipeline developed by \cite{Magro2013}. The GPU beamforming kernel described in 
this paper aims to replace the S-engine by forming the beams within the 
transient detection pipeline, as described in sections \ref{best2Section} and 
\ref{transientSection}.

\section{GPU-Based Beamforming}
\label{beamformingSection}

Real-time beamforming systems pose several challenges. First of all the 
digitised antenna voltage streams need to be transported to backend processors, 
which have to cope with the generally high data rates. Network links and 
devices can achieve very high transfer rates, however in order to offload 
processing to a GPU all this data needs to be transferred through 
Peripheral Component Interconnect Express (PCIe) 
links, which have limited peak bandwidth. This can create an upper bound on the 
number of beams which can be generated and transferred to CPU memory. Also, 
depending on the beam representation scheme, the amount of GPU memory might 
limit the level of parallelism achievable. These factors need to be taken into 
consideration when a beamforming kernel is deployed within a real-time 
pipeline, 
and will be discussed in further detail shortly.

\subsection{Implementation}
\label{implementationSection}

Ignoring the phase-shift calculation method employed to point each beam, the 
computational complexity for a generic coherent multi-beam beamformer is  
$\mathcal{O}(B N_b N_a )$ where $B$ is the bandwidth, $N_b$ is the number of 
synthesised beams and $N_a$ is the number of antennas. The required 
computation rate per input data sample is $\mathcal{O}(N_b)$, suggesting that 
many beams need to be formed to maximise GPU utilisation and efficiency. 
Additionally, there is a fixed upper limit on the data rate into and out of the 
GPU, such that PCIe transfers are rate-limiting. However, if $N_b > N_a$ the 
data rate off the GPU is greater than the rate onto it. Thus, the applicability 
of a GPU beamformer can be determined by $N_a$, $N_b$ and the type of 
processing which occurs after beamforming. For example, transient detection and 
pulsar observations require several compute-intensive processing steps, so for 
such observations a GPU beamformer would be useful for large $N_a$ arrays with 
only a few beams, while space debris observations require limited 
post-processing after beamforming, with data being reduced within the GPU 
itself, such that GPUs can be used for a wider variety of arrays.

The beamforming process is essentially a matrix-vector multiplication, which 
involves multiplying a vector of $N_a$ antennas by an $N_b \times N_a$ matrix 
of coefficients to form $N_b$ beams, per frequency channel. To this end, 
standard CUDA libraries can be used to perform these operations. However, the 
implemented beamformer has to include additional functionality inside the 
kernel which would either have to be performed before and/or after the matrix 
multiply, or by customising existing implementations. These include data 
unpacking (and re-packing if required), absolute power computation, value 
conversion and so on. These extra steps will be discussed shortly. Using a 
separate kernel for these operations would cancel any speed-up gained by using 
readily available libraries due to extra transfer from and to GPU memory. 
The work required to customise existing libraries was deemed to be comparable 
to a custom implementation, and the latter approach was chosen.

Coherent beamforming is trivially paralellisable across the frequency, time and 
beam 
dimensions. A simple implementation would have a single GPU thread 
combine all the antennas, applying appropriate weights, for a single $(f,t,b)$ 
triplet. This would require $2N_a$ global memory requests for every thread, 
since $N_a$ antenna values and $N_a$ complex weights are needed, with an 
instruction count of $8N_a - 1$, resulting in a flop
to data request ratio of 4. Due to the large global memory latencies, such a 
ratio would make this implementation bandwidth limited within the GPU, and thus 
data reuse schemes need to be employed to increase performance. By making the 
assumption that beam coefficients do not need to change within small time 
frames, it is possible to reuse the same weights for all the time bins residing 
in a GPU buffer. This is especially true for wide beams, since it takes 
astronomical objects a relatively long time to pass through the beam, however 
does not hold when very narrow beams are required, or when tracking 
non-astrophysical objects such as satellites. This drastically reduces the 
number of coefficients which need to be generated and read from global memory 
within a thread block. We use this assumption to partition beam generation, 
where a 3D grid is mapped to the output space, with time bins varying in the 
x-dimension, frequency channels along the y-dimension and beam subset along 
the z-dimension. Thus each thread block generates a subset of the beams for a 
number of time bins, for a single frequency channel. Within each thread block, 
local beam accumulators are declared and stored in registers, one per beam, to 
which weighted antenna values are added. The antenna coefficients are loaded 
once per antenna group and stored in shared memory. This antenna partitioning 
is required as otherwise too much shared memory would be utilised, especially 
for large arrays, resulting in a decrease in occupancy and GPU compute resource 
utilisation.

Our kernel was originally implemented to match BEST-2 backend specifications, 
most notably the output data format of the F-engine, which sends out antenna 
signals as 4-bit, two's complement, complex voltages. This 
format enables the packaging of each antenna value as a single byte, and groups 
of 4 antennas can in turn be packaged as 32-bit words. This is beneficial for 
the GPU kernel as each group can be loaded with a single memory request, which 
can be coalesced if they are accessed contiguously by a thread warp. For this 
reason, antennas are processed in groups of 4 in the beamformer, and complex 
coefficients are loaded in groups of the same size. In the inner-most 
loop, where these antennas are accumulated to form beams, this also has the
effect of increasing instruction level parallelism, thus decreasing execution 
time. Breaking down antennas into groups of 4 has the additional benefit of 
making the kernel extensible to larger arrays without affecting performance.

\begin{algorithm}[t!]
 \caption{Coherent beamforming GPU implementation.}
 \begin{algorithmic}
    \STATE Declare $coeffs$ shared memory

    \FOR{time bins to process}
    \STATE \hspace{5mm} Declare local beam accumulators 
$beams[beams\_per\_tb]$

      \hspace{1mm} \FOR{$ant=1$ \TO $nants/4$}

        \STATE Load antenna group from global memory
        \STATE Cooperatively load coefficients subset and store in $coeffs$ 
        \STATE Synchronise threads

        \FOR{$b=1$ \TO $beams\_per\_tb$}
	    \STATE Update local beam accumulator
	\ENDFOR

      \ENDFOR
      \FOR{$b=1$ \TO $beams\_per\_tb$}
	\STATE Store generated beam to global memory
      \ENDFOR 

    \ENDFOR
 \end{algorithmic}
 \label{beamformingAlgorithm}
\end{algorithm}

\subsection{Performance}
\label{performanceSection}

Algorithm \ref{beamformingAlgorithm} provides a detailed breakdown of our 
implementation. The outer time loop is included to handle cases where the 
number of input time bins is not exactly divisible by the total number of 
threads, such that threads will process at most 2 time bins. Greater emphasis 
was put on making the beamforming kernel optimised for NVIDIA Kepler GPUs, 
making the implementation future proof and capable of fully utilising current 
top-range, high performance GPUs. These GPUs have a higher number of registers 
allocated per thread block, however the core clock speed is lower than in Fermi 
GPUs. This means that more data can be kept in fast memory, however care must 
be taken to better hide access latencies and make sure that a high level of 
instruction-level parallelism is achieved. No Kepler-specific instruction were 
utilised, and memory coalescing is handled directly by the kernel without the 
assumption of a GPU cache, such that the kernel is able to run efficiently 
on any GPU architecture by optimally tweaking configuration parameters.Our 
implementation requires two configuration parameters to be defined and 
consequently optimised upon:

\begin{description}
 \item {\bf Number of accumulators} defines the number of beams each thread 
will generate. These are stored in registers and thus are limited by the number 
of registers available to a thread block. A lower number of accumulators will 
decrease weight values reuse and increase global memory bandwidth requirements, 
while a high number will reduce GPU occupancy and in the worst case can be 
spilled to local memory, significantly reducing performance.
 \\
 
 \item {\bf Threads per block} determines the number of time samples a single 
thread block will process (one per thread). A low number will reduce the number 
of active warps in a SMX, leading to diminished parallelism, while a higher 
number will increase shared memory and register requirements, reducing 
occupancy.
\end{description}

\begin{figure}
  \centering
  \includegraphics[width=225pt]{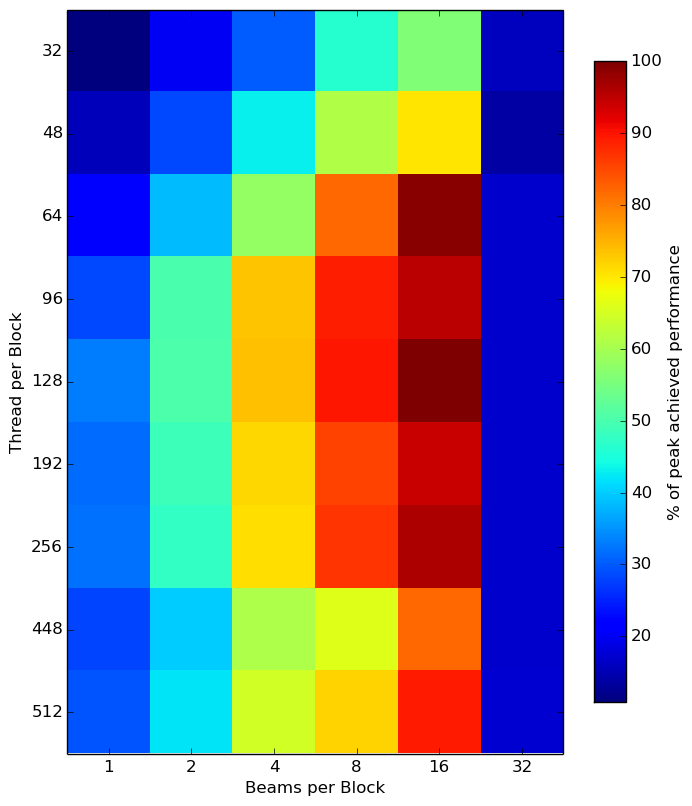}
  \caption{The result of configuration optimisation runs for the 
	   beamforming kernel, where 128 threads per block, each generating 
	   16 beams, provides the best parameter combination on the 
	   test device.}
  \label{gpuBeamConfig}
\end{figure}

In order to find the optimal configuration for the above parameters a series of 
benchmark tests were conducted on an NVIDIA Tesla K20 card for a range of 
parameter 
combinations. Figure \ref{gpuBeamConfig} shows the optimal values on 
the test device, where 16 accumulators and 128 threads per block provide the 
best performance. For these tests a 420 ms simulated buffer containing 32 4-bit 
complex sampled antenna voltage streams with a 20 MHz bandwidth channelised 
into 1024 frequency channels was generated and processed 10 times for each 
combination. The resulting mean was used as a measure of configuration 
performance. For these and subsequent tests we used Compute Unified Device 
Architecture (CUDA) 6.0 (CUDA compilation 
tools release 5.5, V5.5.0).

An in-depth analysis of the algorithm's performance was also performed, 
a summary of which is presented in Table \ref{gpuPerformanceTable}. These value 
were generated by the NVIDIA CUDA profiler, for which the highest performing 
configuration for BEST-2 parameters was used. The beamformer's compute and 
memory utilisation is balanced, with $\sim$80\% of compute and $\sim$60\% of 
L1/Shared memory being utilised. The kernel is compute-bound, relying heavily 
on the use of shared memory to cache beamforming weights, achieving a shared 
memory bandwidth of 1.3 TB/s with a shared memory efficiency of 50\%. A high 
level of instruction level parallelism was also achieved, at 4.5 instructions 
per cycle. 
The FLOP rate achieved for the highest performance configuration (16 
accumulators per threads and 128 threads per block) is 1.3 TFLOPS, 
approximately 38\% of the peak theoretical performance of the Tesla K20. The 
number of FLOPS achieved is calculated by counting the number of floating point 
operations required for each complex multiply, which in this case of 8$N_a$ per 
beam/channel/timestep. 

\begin{table}
  \centering
  \begin{tabular}{  r  l  }
    \hline
    \multicolumn{2}{ c }{Test Device: NVIDIA Tesla K20c} \\
    \hline
    Kernel compute utilisation          & $\sim$80\% \\
    Warp efficiency                     & 100\% \\
    Shared memory throughput            & 1,289 GB/s    \\
    Major issue stall                   & Execution dependency (50\%)  \\
    Issued instructions per cycle       & 4.489   \\
    Shared memory efficiency            & 50.1\% \\
    FLOPS                               & 1.3 TFLOPS \\
    \hline
  \end{tabular}
  \caption{GPU beamforming kernel performance analysis.}
  \label{gpuPerformanceTable}
\end{table}

\begin{figure}
\begin{center}
\includegraphics[width=220pt]{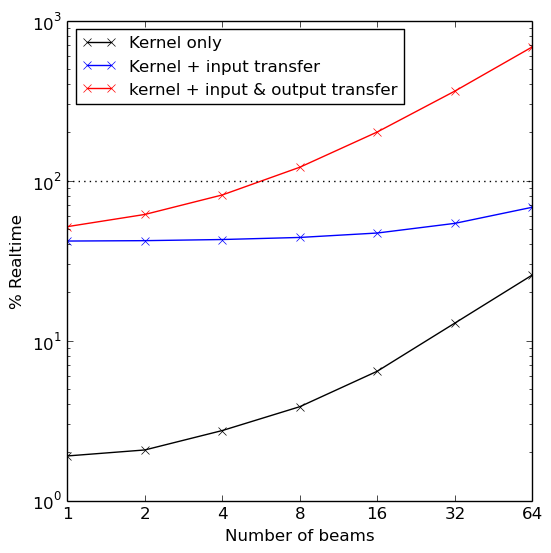}
\end{center}
\caption{Coherent beamforming performance benchmarks for varying number of 
synthesized beams, for BEST-2 parameters.}
\label{beamformerParameter}
\end{figure}

A scalability analysis of the algorithm was also performed, where a 420 ms data 
buffer containing voltage data from 32 single polarisation antennas with a 20 
MHz band channelised into 1024 frequency channels was used to generate an 
increasing number of synthesised beams in one GPU iteration. Figure 
\ref{beamformerParameter} shows that performance scales linearly with 
increasing number of beams and that execution time is a fraction of real-time 
for BEST-2 parameters. The maximum number of output beams which can be 
optimally synthesised is determined by the amount of global memory available on 
the GPU. The GPU spends 50\% of its time waiting for raw voltage data to be 
copied from host to GPU memory, thus indicating that the implementation is 
bandwidth limited over the PCIe link. When excluding the 
time required to copy 
the generated beams out of GPU memory the total execution time is evenly split 
between kernel execution and transferring antenna voltages to GPU memory. This 
data overhead can essentially be masked by using two CUDA streams if the 
pipeline is capable of overlapping kernel execution and data transfer for the 
following iteration, however this would also require additional global memory. 
The situation is worse when the generated beams are post-processed externally 
to the GPU on which they were generated, as in this case the pipeline will be 
dominated by PCIe transfer overheads. This can also be alleviated somewhat by 
increasing the number of streams on GPUs with dual copy-engines, in which case 
the total execution time per iteration is the duration of the slowest stage 
(PCIe links are full-duplex, so the PCIe transfer time is the time required to 
transfer the largest buffer to or from GPU memory).

This poses several challenges for GPU-based beamforming pipelines and raises 
doubts on whether GPUs are a viable platform for large-N aperture array 
telescopes. Processing the generated beams on the GPU itself increases the 
compute-to-copy ratio. Additional processing can take the form of correlation, 
accumulation for image generation or transient detection. The latter 
alternative will be discussed in further detail in Section 
\ref{transientSection}.

\subsection{CPU Comparison}

In order to measure the performance speed-up gained by using GPUs, we 
also implemented a CPU version of the same algorithm, with each antenna voltage 
sample packed into a single byte (4-bits per complex component). The code was 
compiled with the Intel icc compiler version 14.0.3. The CPU beamformer was 
implemented with SIMD functionality in mind, making sure that the compiler can 
perform optimised auto-vectorisation where required. Appropriate benchmarks 
were 
also performed. These were run on a system with two Intel Xeon E5-2640 CPUs 
running at 2.5 GHz (can be Turbo-boosted to 3.0 GHz) with hyper-threading 
disabled, resulting in a total of 12 physical cores. OpenMP was used to 
parallelise the implementation across multiple cores. The host operating system 
was Ubuntu 
14.04 with Linux kernel 3.13.

The peak theoretical performance of a CPU can be computed using 
$N_{\text{fpc}} \cdot N_{\text{cores}} \cdot N_{\text{freq}}$, where 
$N_{\text{fpc}}$ is the number of floating point operations per clock cycle,  
$N_{\text{cores}}$ is the number of physical CPU cores and $N_{\text{freq}}$ is 
the maximum processor frequency. For the Xeon E5-2640, this results in a peak 
performance of 120 GFLOPS (six cores at 2.5 GHz and 8 FLOPS per cycle). 

\begin{figure}
\begin{center}
\includegraphics[width=220pt]{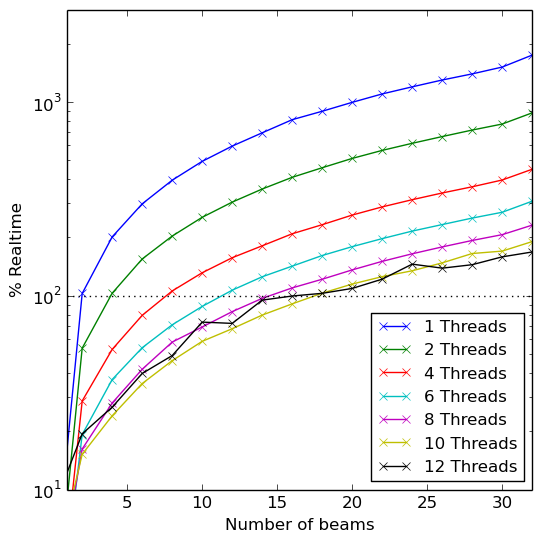}
\end{center}
\caption{CPU beamformer performance benchmarks when synthesising a range of 
beams using varying number of threads, for BEST-2 parameters.}
\label{beamformerCPUParameter}
\end{figure}

On this test system the CPU implementation achieves 0.45 cycles per 
instruction, where the Xeon E5-2640 can issue up to four instructions per 
cycle, with a theoretical peak of 0.25, suggesting that a high level of 
instruction level parallelism and SIMD functionality was achieved. Figure 
\ref{beamformerCPUParameter} shows the result of these benchmarks, where a 
range of beams were synthesised using varying number of threads for BEST-2 
parameters. Performance increases linearly when using up to 10 threads, in 
which case up to 16 beams can be generated in real-time. With this 
configuration, each CPU (6 threads) achieves the equivalent performance of 90 
GFLOPs (assuming 8 FLOPs per complex multiplication), suggesting that 75\% of 
the CPU is being utilised. 

\begin{figure}
\begin{center}
\includegraphics[width=225pt]{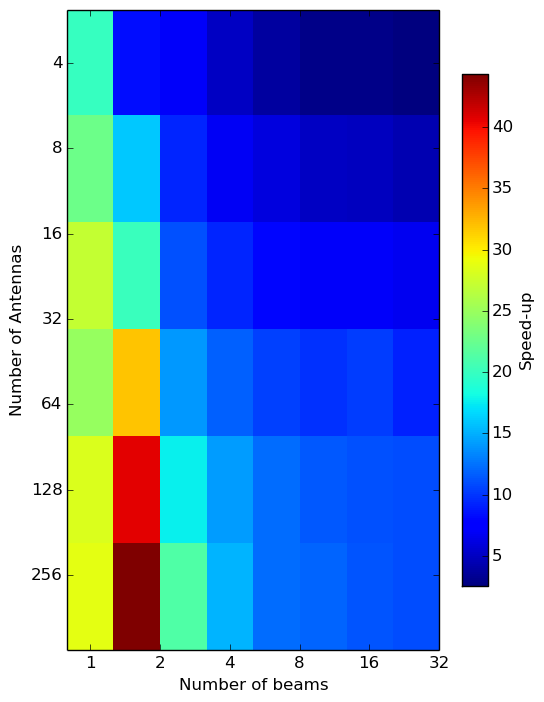}
\end{center}
\caption{GPU (NVIDIA K20) speedup when compared to an optimised CPU (dual Xeon 
E5-2640) implementation.}
\label{beamformerSpeedup}
\end{figure}

Figure \ref{beamformerSpeedup} shows the speed-up gained when using the GPU 
beamformer compared with the CPU beamformer. When the number of antennas is 
small and a large number of beams are required, speed-up is minimal, 
whilst speed-up is maximised for large-arrays and a small number of synthesised 
beams. For BEST-2 parameters, the overall speed-up is of an order of magnitude 
($\sim10z$).

\subsection{Comparison to Related Work}

GPU-based beamforming is a relatively unexplored area in radio astronomy, 
possibly due to the assumption that any such system will be severely limited by 
the PCIe bandwidth required to transfer data to the GPU. \cite{Nilsen2009} 
presented two implementations for digital beamforming using CUDA and 
benchmarked on an NVIDIA GeForce 8800. They state that GPUs are a suitable 
platform for high-performance beamforming. \cite{Sclocco2012} have implemented 
a 
GPU-based version of the Blue Gene/P beamformer within the LOFAR pipeline, and 
state that they achieve a 45-50 times speed up, at 642 GFLOPS when using CUDA, 
on an NVIDIA GTX 580 (using 40\% of the peak theoretical performance), with a 
power efficiency improvement of 2-8 times. This algorithm was developed and 
benchmarked on Fermi-architecture GPUs, so directly comparing it with our 
implementation, which is optimised for the Kepler architecture, does not 
provide an adequate comparative metric. Both implementations achieve about 40\% 
of peak performance, however our kernel can sustain this utilisation rate over 
a wide parameter range, whilst the one presented by \cite{Sclocco2012} is 
optimal for particular values. On the other hand, if an odd number of beams 
or antennas is required, our implementation will suffer a degradation in 
performance. Both implementations are targeted for a specific use case (LOFAR 
and BEST-2), and their applicability to other instruments has to be 
investigated.

\cite{Magro2014} discuss the possibility of generating all SKA1-Low station 
beams and performing first stage channelisation on GPUs, however the required
PCIe bandwidth ends up being a severe bottleneck for a GPU-based architecture, 
resulting in approximately 16 GPUs per station.

\section{Deployment at the BEST-2 Array}
\label{best2DeploymentSection}

The beamforming kernel was implemented and optimised to process the coarsely 
channelised, complex data output by the F-engine deployed at the BEST-2 
array. This data is streamed out as User Datagram Protocol (UDP) packets, with 
UDP headers encapsulating 
groups of 128 time samples from 32 antennas, for each frequency channel, thus 
resulting in 4 KB packets with an additional 64-bit header containing the 
timestamp of the first time bin in the packet as well as the frequency channel 
index. The total output data rate can be calculated using $D=C\times T \times A 
\times W$, where $C$ is the number of frequency channels, $T$ is the number of 
samples per second, $A$ is the number of array elements and $W$ is the word 
length. In our case, $C$ = 1024, $T$ = 19531.25, $A$ = 32 and $W$ = 8-bits 
(4-bits for each complex component), resulting in a total output bandwidth of 
5.12 Gbps, excluding packet headers. Therefore a single 10 GigE link is 
sufficient for data transfer between the F-engine and GPU server. At the 
receiving end, the packet receiver used for \cite{Magro2013} was updated to be 
compatible with this format. Groups of 128 time samples are considered as SPEAD 
heaps, and thus heap functionality is still applicable. An additional lookup 
table is required to match the antenna order sent by the F-engine to the 
element 
ordering within the array and antenna configuration file, and is performed in 
the buffering thread. The kernel was integrated with two software pipelines.

\subsection{Beamforming Pipeline}
\label{beamPipelineSection}

A beamforming pipeline was implemented and deployed at the BEST-2 array, 
using the beamforming kernel described in section \ref{implementationSection}. 
This pipeline is currently being used for space debris detection prototyping 
and test observations \cite{Morselli2014}. A future paper will discuss the full 
implementation of this system. Several test observations were conducted during 
deployment. For these tests, additional functionality was included into the 
pipeline, including the ability to generate an arbitrary number of stationary 
(within the primary field of view) and tracking (fixed on a celestial object) 
beams, as well as sub-arraying. The latter uses an antenna mask in the kernel 
to only include voltages for the required antennas. Additionally, a conversion 
mechanism was included to change the sample representation mechanism employed 
by the digital back-end to one which better suits the GPU. The kernel 
performance impact of these additional computations features since they are not 
performed in the inner most loop, and conversions use a lookup table stored in 
constant memory. 

\begin{figure}
  \begin{center}
  \includegraphics[width=260pt]{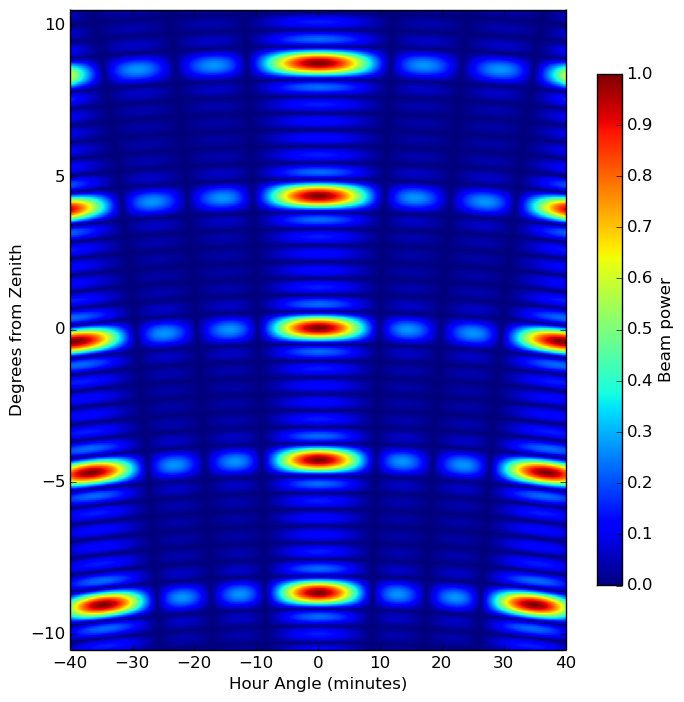}
  \end{center}
  \caption{Simulated beam pattern for the BEST-2 array.}
  \label{beamPattern}
\end{figure}

\begin{figure*}
  \begin{center}
  \includegraphics[width=320pt]{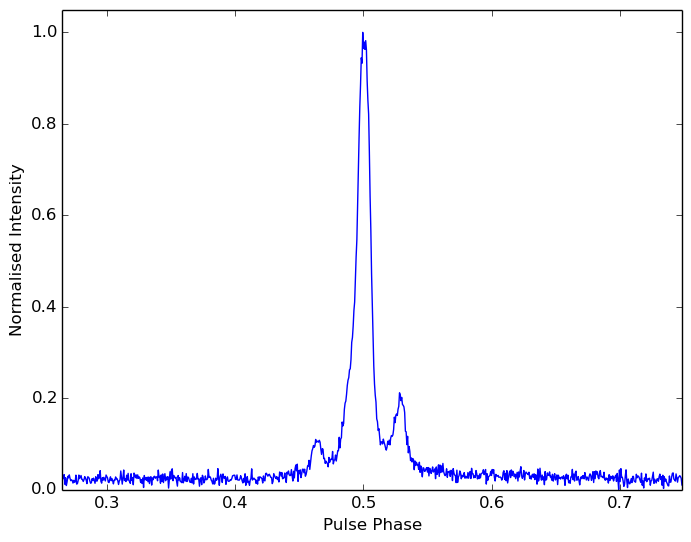}
  \end{center}
  \caption{Integrated pulse profile of pulsar PSR B0329+54 consisting of 200 
profiles.}
  \label{pulsarFigure}
\end{figure*}

\begin{figure*}
  \begin{center}
  \includegraphics[width=400pt]{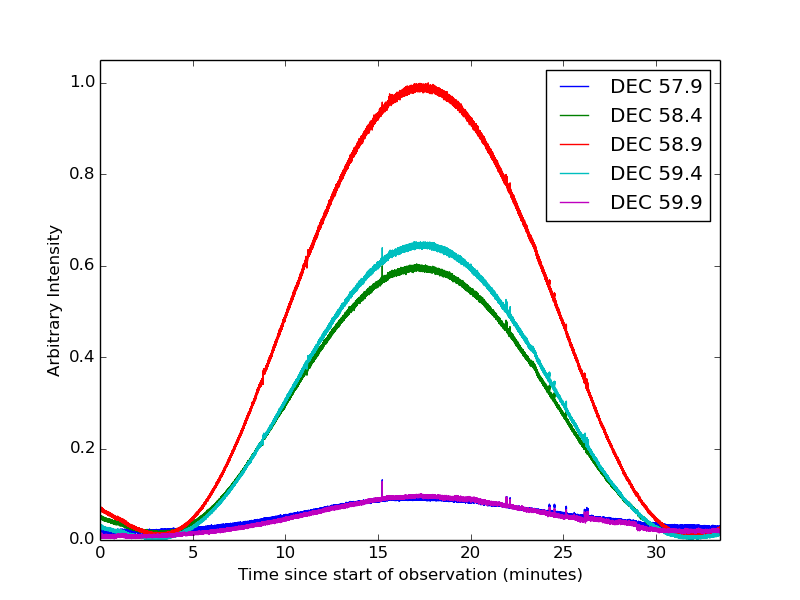}
  \end{center}
  \caption{Multi-beam observation of Cassiopeia A transiting across multiple 
synthesised beams at varying declinations.}
  \label{casaFigure}
\end{figure*}

Figure \ref{beamPattern} shows a simulated beam pattern of the 
BEST-2 array. This was performed by generating the beamforming coefficients for 
a range of target RAs and DECs and computing the power of the beam at each 
point. About 32 beams are required to tile the entire primary beam with 
synthesised beams. The number of GPUs required to perform 
all the processing depends on the what processing stages are required after 
beamforming. For examples, a bi-static space debris detection system can run on 
a single GPU for 32 beams, while a transient detection pipeline will require 
considerably more GPUs. 

Figure \ref{pulsarFigure} shows the result of a single beam test 
observation of pulsar PSR B0329+54. The beamformed data was stored to disk, 
with dedispersion and folding performed offline. In this case, the profile was 
folded 200 times. A tracking beam was used, fixed to the pulsar's RA and DEC, 
and the data used for generating the profile coincides with the pulsar being in 
the center of the primary beam. Figure \ref{casaFigure} shows the result of a 
multi-beam observation of Cassiopeia A, with stationary beams pointed at an 
hour angle of 0 and each having an offset which is a multiple of 0.5 DECs from 
the source's DEC. The $\sim$34 minute window of the observation is plotted, 
coinciding with the source transiting across the primary beam.

\subsection{Transient Detection Pipeline}
\label{transientSection}

We have also integrated the GPU beamforming kernel into the GPU-based transient 
detection pipeline described in \cite{Magro2013}. This section provides general 
design considerations for creating a transient detection pipeline with 
beamforming capabilities, assuming that a single GPU is capable of running the 
transient detection processing stages for at least one synthesised beam. The 
total number of GPUs required will depend on the surveying and beamforming 
requirements for the telescope and science case on which the pipeline is 
deployed. 

The beamforming kernel has to be executed before the RFI excision and 
dedispersion stage, after which each beam is then processed by a separate GPU 
instance. The main challenge lies in determining the most efficient way to 
generate these beams, minimising data movement between the host and GPUs, as 
well as amongst GPUs. Three schemes can be employed.

The simplest scheme, implementation wise, would be to have each GPU 
processing thread generate its own coherent beam. This would require the input 
data buffer to be copied to each GPU instance, thus replicating this buffer 
multiple times within a GPU, greatly reducing the number of time spectra which 
would fit in global memory, as well as resulting in a degradation in 
performance due to the time spent transferring data over PCIe links. Also, the 
beamforming kernel is optimised for generating multiple beams in parallel, and 
this scheme would not be fully utilising the kernel's performance capabilities

At the other extreme, a single GPU can generate all the beams required 
by all the processing threads and then copy each beam to its destination, 
which could be on a separate GPU. Only one host to GPU transfer is required, 
and kernel execution time is minimised when compared to the first scheme where 
multiple kernel launches are required, each generating one beam, which is 
clearly inefficient. However, a copy per generated beam is required, some of 
which can be across GPUs for multi-GPU systems. This scheme also introduces 
heterogeneity across GPUs in the pipeline, where one GPU acts as a producer 
which provides consumers with data to work on. 

An alternative approach is to combine both schemes, where beams are generated 
in the same kernel launch on the GPU on which further processing will be 
performed. The input antenna voltages need to be copied once to every GPU, 
where a 'master' thread launches the beamforming kernel, generating all the 
required beams. Upon completion each processing thread is provided with a 
pointer to its input beam, and processing advances as per standard transient 
detection.

\begin{figure*}
  \begin{center}
  \includegraphics[width=350pt]{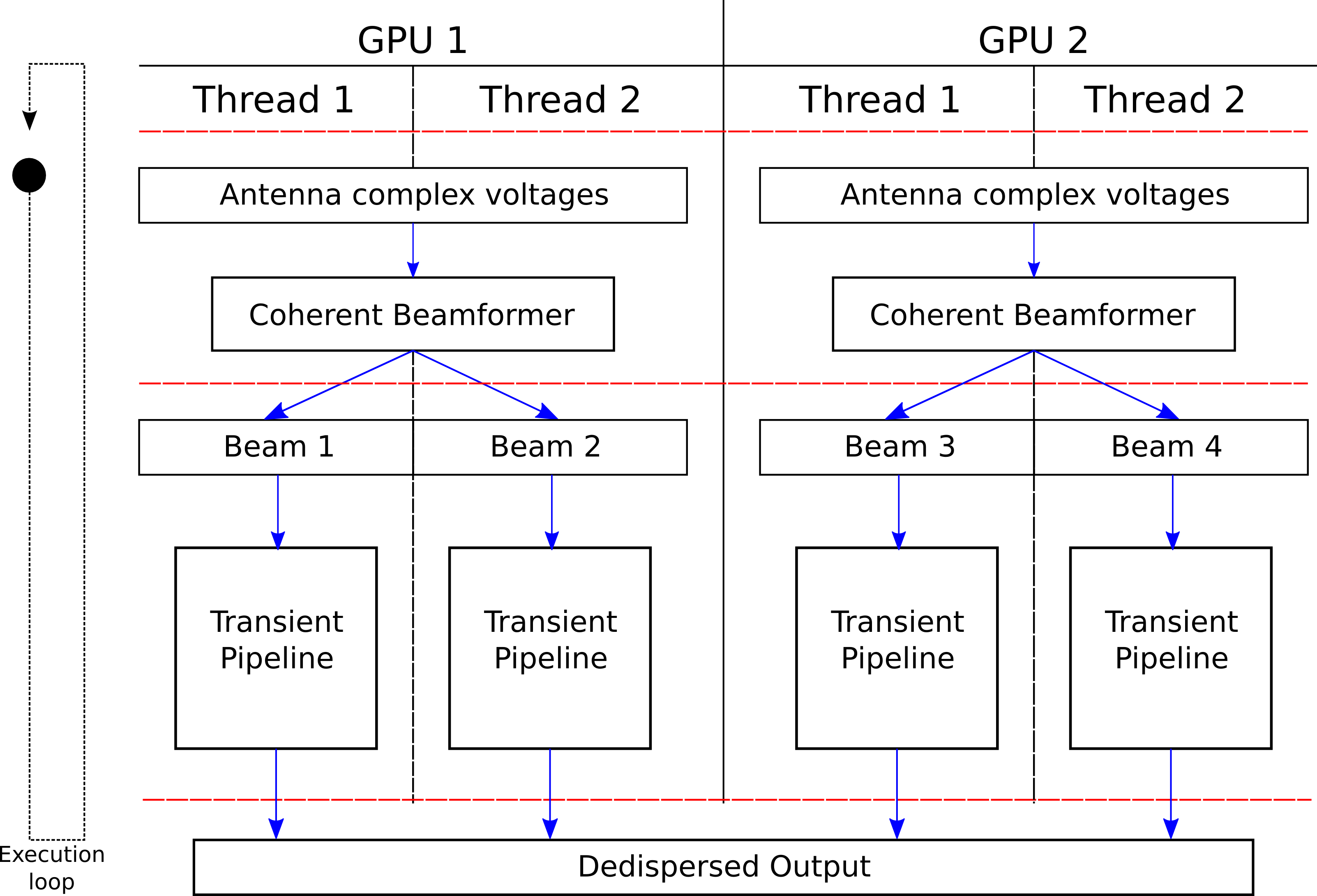}
  \end{center}
  \caption{A schematic of the transient detection pipeline with integrated 
coherent beamforming kernel.}
  \label{beamformerPipeline}
\end{figure*}

The latter scheme provides a good balance between input data replication, data 
transfer over PCIe links and beamforming kernel efficiency, and is the scheme 
we adopted for the transient pipeline. The resulting design is illustrated in 
Figure \ref{beamformerPipeline}. The buffered antenna voltages are copied to an 
input buffer allocated on every attached GPU. These transfers occur in parallel 
over different hardware PCIe links. A CPU thread per beam is created, which is 
responsible for all GPU-related processing of its associated beam. Since 
multiple beams can reside on the same GPU, one of the CPU threads is 
initialised as the master thread associated with the GPU (belonging to the 
first allocated beam). This is responsible for generating all the GPU buffers 
and launching the beamforming kernel. It will generate all the beams in a 
shared output buffer on the GPU, which can be accessed directly from other 
CPU threads mapped on the same GPU, thus not requiring a copy per generated 
beam. The transient detection pipeline is then launched for each synthesised 
beam, keeping the beamformed data in GPU memory. The output of the transient 
detection pipeline is then processed on the CPU, as described in 
\cite{Magro2013}. 

The main shortcoming of this design is that the 'slave' processing threads have 
to wait for the beamforming kernel in the master thread to finish before they 
can start processing data, thus wasting valuable clock cycles. This duration is 
relatively short when compared to dedispersion. In the current implementation, 
the slave threads are blocked in a barrier synchroniser. The beamforming 
coefficients are computed by the pipeline manager whilst waiting for the 
processing threads to finish working on the previous iteration's input buffer. 
This allows beam pointings to change dynamically during pipeline execution, 
which is useful for tracking observations or online follow-up of interesting 
transient events, provided appropriate feedback mechanisms exist within the 
pipeline itself. The beamforming weight computation overhead is negligible 
relative to the execution time of a single pipeline iteration. The array 
configuration is stored in an XML file containing relative antenna locations. 

This pipeline was benchmarked using BEST-2 parameters as a reference 
telescope. A 2.52s buffer containing voltage data from 32 20 MHz single 
polarisation antennas centered at 408 MHz, split into 1024 frequency channels, 
was generated and processed in a single pipeline iteration. One NVIDIA K20 card 
was used for this test. Four beams were synthesised from this buffer, each of 
which was then dedispersed over a range of 864 Dispersion Measure (DM) trials 
with a maximum DM of 
86.4 pc cm$^{-3}$. Table \ref{timingTable} lists the execution time 
of each stage. The CPU to GPU transfer of antenna voltages and the beamforming 
kernel launch are performed by the 'master' processing thread, and the input 
data buffer is limited by the amount of global memory available. This table 
shows that, for a 32 element array, the beamforming cost is negligible when 
compared to the total execution time for the entire pipeline, and amounts to 
approximately 3\%. This hints to the possibility of deploying a fully GPU-based 
beamforming and transient detection system for small, low-bandwidth arrays. 
Analogue signal reception, digitisation and equilisation still need to be 
performed prior to beamforming, and these operations are more suitable to an 
FPGA-based system, such as the F-engine deployed at the BEST-2 array.

\begin{table}
  \centering
  \begin{tabular}{ p{3.5cm}  l }
    \hline
    \multicolumn{2}{c}{Antenna Voltage Copy Time: 291.92 ms} \\
    \hline
    Beamforming      & 74.98 ms \\ 
    Bandpass Fitting & 80.16 ms \\
    RFI Thresholding & 36.44 ms \\
    Dedispersion     & 1780.60 ms \\
    Median Filtering & 70.96 ms \\
    Detrending       & 44.16 ms \\
    \hline
    \multicolumn{2}{c}{Copy from GPU: 101.64 ms} \\
    \hline
    \multicolumn{2}{c}{Total iteration time: 2379.27 ms} \\
    \hline
  \end{tabular}
  \caption{GPU timings for one pipeline iteration performing beamforming, 
bandpass-fitting, RFI thresholding, dedipsersion, median filtering and 
detrending. }
  \label{timingTable}
\end{table}

\section{Conclusion}

We have presented a CUDA-based GPU implementation of a coherent beamformer 
specifically designed and optimised for deployment at the BEST-2 array, 
connected to the digital backend developed by \cite{Foster2014}. This 
beamformer can generate an arbitrary number of synthesized beams for a wide 
range of parameters. With optimal kernel parametrisation, the kernel achieves 
1.3 TFLOPs on an NVIDIA Tesla K20, using $\sim$80\% of available GPU compute 
resources. Comparing this implementation with a similarly optimised CPU version 
of the same algorithm run on a dual Intel E5-2640 system shows an overall 
performance gain of an order of magnitude ($\sim$10x). We have integrated this 
kernel with two real-time software pipelines at the BEST-2 array, a standalone 
beamforming pipeline and the transient detection pipeline developed by 
\cite{Magro2013}, and show that when the beamformer is used in conjunction with 
other processing kernels, and as long as the generated beams stay in GPU 
memory, beam generation overhead is minimal when compared to processes such as 
dedispersion. This can alleviate the PCIe data movement bottleneck for 
specialised pipelines. 

\section*{Acknowledgements}
\label{ack}

The Croce del Nord  Radio Telescope is a facility of the University of Bologna 
operated under agreement by the INAF Istituto di Radioastronomia. 

We would like to thank Stelio Montebugnoli, Jader Monari, Germano Bianchi, 
Andrea Mattana, Giovanni Naldi and all 
the staff at the Medicina Radio Observatory for their invaluable help on-site 
during deployment and observing sessions. We would also like to thank the 
system designers for the digital backend, especially Griffin Foster, for their
help in designing the interfacing protocol between the digital backend and host 
system, as well as for hours of support.

\end{document}